# Application of Nash equilibrium for developing an optimal forest harvesting strategy in Toruń Forest District

Jan Kotlarz[a*]

[a] Łukasiewicz Research Network– Institute of Aviation , Al. Krakowska 110/112, 02-256 Warsaw, Poland
[*] Corresponding author: e-mail: jan.kotlarz@ilot.lukasiewicz.gov.pl

## Abstract

This study investigates the application of Nash equilibrium strategies in optimizing forest harvesting decisions, focusing on multiple management objectives in forestry. Through simulation-based analysis, the research explores the evolution of various indicators during the game: 1) the mass of $CO_2$ sequestration, 2) forest stands biodiversity, 3) the harvested wood volume, 4) native species fraction, and 5) protective functions. The results underscore the importance of considering diverse objectives and balancing competing interests in forestry decision processes. The forest stands designated for harvesting in the Toruń Forest District were defined as the initial strategy, and indicators for all objectives were calculated accordingly. A Nash equilibrium was identified through a game involving five players representing individual objectives with partially conflicting aims. The final strategy was obtained by modifying specific forest stands designated for harvesting, thereby maintaining the planned wood volume extraction while simultaneously reducing biodiversity loss by nearly 40%, preserving protective functions across over 600 hectares of forested areas, enhancing decadal carbon sequestration in the forest district by 100,000 tons, and additionally improving species suitability by nearly 10%. The findings suggest the potential for further research and refinement of Nash equilibrium-based optimization approaches to enhance the effectiveness and sustainability of forest management practices.

## 1. Introduction

Nash equilibria have long been the subject of research within forestry economics. They are applied in issues such as conflicts arising from competing forest use among stakeholder groups with opposing objectives (Shahi & Kant, 2007), competition for limited forest resources (Lamaei, 2010), and even in seeking equilibria in games with seemingly zero-sum outcomes between the management of the US National Forests and environmental movements (Carlson & Wilson, 2004). As a result of adopting Nash equilibria in such games, all market and societal players have an interest in maintaining this state, as no player can improve their outcome through unilateral changes in strategy. In Nash equilibrium, each player makes the best possible decision considering the strategies of other players, leading to stability in the situation (van Damme, 1991; Badev, 2021).

In many cases, making decisions as a public forest manager is also a non-trivial issue, as contemporary sustainable management imposes on this market player a set of partially conflicting and partially convergent objectives. The largest forest manager in Poland is the State Forests National Forest Holding. According to Article 7 of the Polish Forestry Law, forest management is conducted according to a forest management plan, taking into account "sustainable forest management" aimed at utilizing forests in a way that preserves their biological richness, maintains high, stable, and predictable productivity over time, and ensures regenerative potential (Rakoczy 2010). The achievement of this policy is measured in forest management plans by the degree of realization of five criteria described i.e. in the documents of State Forests National Forest Holding (Pacia 2021).

- **Criterion 1:** Conservation and appropriate enhancement of forest resources and their contribution to the global carbon balance. Forest management should maximize carbon accumulation from the atmosphere. Carbon accumulation decreases as stands age due to reduced photosynthetic capacity. After reaching a specified age in the plan, harvesting processes should be initiated (Asshof et al., 2006). Carbon accumulation is measured in tons of $CO_2$ per hectare per year.
- **Criterion 2:** Maintaining the health and vitality of forest ecosystems. According to this criterion, the processes carried out should make the most of natural structures and processes. At the same time, it is important to support species and structural diversity in individual habitats. The diversity index can be measured using the Shannon-Wiener index (Kotlarz et al., 2018).
- **Criterion 3:** Maintenance and enhancement of the productive functions of the forest. This criterion requires maximizing the extraction of forest products within limited dimensions while maintaining this capacity over the long term. The level of attainment is also constrained by minimizing negative environmental impacts. An indicator of this criterion may be the amount of harvested timber per year measured in $m^3$/year.
- **Criterion 4:** Conservation, enhancement, protection of biodiversity. This criterion implies a preference for: natural regeneration, native species, and natural ones consistent with habitat conditions, processes that differentiate species composition and vertical and horizontal habitat structure. An indicator of this criterion may be the percentage conformity of cultivated species with the habitat.
- **Criterion 5:** Conservation and enhancement of the protective functions of the forest. This criterion primarily concerns the protection of soils and water in forest compartments w wysłałeśith protective functions and in reserves, especially in riparian, swampy habitats, and moist forests. The indicator for this criterion can be the number of forest stands with protective characteristics.

The simultaneous fulfillment of all criteria is only possible to a certain extent, as maximizing timber extraction, for example, may decrease species diversity and, in many cases, reduce carbon dioxide accumulation from the atmosphere. On the other hand, actions aimed at achieving individual criteria may be compatible with each other. For instance, to increase carbon dioxide accumulation in a forest stand over 120 years old, it may be necessary to remove some trees and replace them with young plantations. This is because each tree species has a negative carbon dioxide balance only during the period of wood mass growth. Therefore, it can be concluded that making decisions to implement specific procedures in forest management planning is not straightforward in terms of maximizing all indicators within the Forest District.

In this study, a negotiation model was developed in which five players were defined within the Forest District, one for each criterion. Each player's goal is to maximize the indicator corresponding to their criterion. The aim is to investigate the utility of Nash negotiation strategies. For simulation purposes, forest plantations located in Poland within the Toruń Forest District were selected.

## 2. Materials and Methods

The data used in this study is based on the Forest Data Bank, which undergoes regular updates on an annual basis through on-site forest inventories conducted by forestry authorities (Talarczyk 2015). Each entry in the dataset provides details about an individual forest stand, defined in forestry as a uniform forest area distinguished by economically significant characteristics that necessitate consistent management practices. These individual forest stands typically cover areas ranging from one to several hectares. The database for the year 2023 for the Toruń Forest District encompassed a total of 8 860 records. For the subsequent analysis, specific forest stand parameters were considered, including

(1) forest stand habitat (i.e. deciduous forest, coniferous forest, dry coniferous forest, alder forest and so on),

(2) forest atand area in hectares,

(3) trees species present in the forest stand with the percentage resolution 10% (0%, 10%, 20%...),

(4) trees species age,

(5) timber volume $v$ for vegetation seasons for each species in each forest stand.

Example records data is shown in the Table 1.

**Table 1**. Example record based on Forest Data Bank.

| Parameter | | Value |
|---|---|---|
| **Forest address** | | 12-24-1-02-109  -f |
| **Habitat** | | Fresh deciduous forest |
| **Area** | | 3,25 ha |
| **Species: Pine** | **Cover** | 70% |
| | **Age** | 59 |
| | **Standing wood volume** | 212 m$^3$/ha |
| **Species: Birch** | **Cover** | 30% |
| | **Age** | 59 |
| | **Standing wood volume** | 66m$^3$/ha |

### 2.1. Prediction of carbon dioxide accumulation

Carbon dioxide accumulation in the atmosphere depends primarily on three factors:

1) current wood density in the plantation,
2) species, and
3) age of trees.

To estimate $CO_2$ accumulation, reference wood growth models for individual species and for individual ages ranging from 20 to 120 years were utilized as calculated in Kotlarz and Bejger (2024). For the age range of 0 to 20 years, the models were extrapolated based on data from the age of 20 to 40 years with the condition that the growth for age 0 is 0 m$^3$ ha$^{-1}$. Approximately 50% of the dry wood mass consists of carbon extracted from the atmosphere. It is assumed that for individual species, 0.45 to 0.55 of the growth mass consists of carbon extracted from the atmosphere. For each record, the predicted total amount of $CO_2$ absorption in tons for the period 2023-2033 was calculated. In the case of determining harvesting for this forest compartment, this value represents lost carbon accumulation. In the case of leaving this forest stand, the value of lost carbon accumulation is zero. The total lost carbon accumulation for all records is denoted as $X_1$.

## 2.2. Species diversity

To calculate the species diversity of forest plantations, the Shannon-Wiener index was applied: $SW = -\sum p \log_{10} p$ where $p$ is the percentage share of the species in the forest plantation, and the summation is performed over all species. In the case of a record in Table 1, SW = - 0.7 $\log_{10}$ 0.7 - 0.3 $\log_{10}$ 0.3 = 0.7 * 0.155 + 0.3 * 0.523 = 0.265. This index for a monoculture of any species is 0, and for maximum diversity (10 species with 10% share in the plantation each) it is 1. The index was calculated for each record and the average was calculated for records designated for harvesting, denoting it as $X_2$. Harvesting a monoculture does not increase $X_2$; the higher the species diversity of the harvested forest stand, the more increased the value of $X_2$.

## 2.3. Timber volume

In this publication, the harvesting of entire forest plantations is analyzed. Therefore, for each record $n$, the volume of standing forest for all species was summed and denoted as $v$. The sum of $v$ for all records designated for harvesting was designated as $X_3$:

$$X_3 = \sum v_n$$

## 2.4. Native species

Based on the Forest Management Instructions, the corresponding tree species were identified for each forest habitat. For example, for the record in Table 1, the habitat is "fresh deciduous forest." According to the Instructions, the appropriate species for this habitat are oak, beech, spruce, and fir. The percentage share of these species in the forest stands is therefore 0%. For each record designated for harvesting, this parameter was calculated, and then their average was calculated as $X_4$. A value of $X_4$ close to zero indicates that forest stands not corresponding to native species were selected for harvesting, while a value close to 1 indicates the selection of stands corresponding to native species.

## 2.5. Protective functions

For the crops designated for harvesting, those with protective functions were summed. The sum of these was placed as the value $X_5$.

## 2.6. Definition of players and payoff values in the payoff matrix

In the decision-making process regarding the selection of forest stands for harvesting, we encounter several constraints. We would prefer to remove stands for which the carbon dioxide accumulation is lower compared to others, those with lower species diversity, those with higher standing timber volume, those with improper tree species in terms of habitat, and those lacking protective functions. Unfortunately, these objectives often conflict with each other. For example, for the record shown in Table 1, due to species habitat mismatch, this stand should be removed. However, the dominant age of pine indicates significant potential for carbon dioxide accumulation in this stand (pine harvest age is approximately 100 years).

Selecting suitable stands for harvesting is defined as a non-zero-sum game. It is assumed that five players participate in the game, one one for each criterion. The first player will seek to minimize lost carbon accumulation, represented by the value $X_1$. The second player will aim to select stands with low species diversity, thereby minimizing $X_2$. The third player will strive to maintain the planned wood volume harvesting plan $X_3$. In this game, it was assumed that strategies harvesting a volume equal to or higher than the harvesting plan have the same utility for player 3 (who does not aim to maximize volume extraction).The fourth player will aim to designate stands that are most incompatible with the habitat, hence minimizing $X_4$. The fifth player will seek to minimize the number of stands with protective functions, thereby minimizing $X_5$.

For the set of stands designated for harvesting, values ($-X_1$, $-X_2$, $X_3$, $-X_4$, $-X_5$) can be calculated and assumed to be the payoffs for each player.

### 2.7. Course of the game and application of Nash equilibria

In this study, it was assumed that the initial scenario, represented by the initial set of stands designated for harvesting ($H_0$), corresponds to those actually designated in the forest management plan for the year 2023. In the simulation beginning forest stands designated for harvesting ($H$) were set as $H_0$. Subsequently, random stands were selected as follows:

a) one from the current set ($a \in H$),

b) ten stands ($b_1$, $b_2$, ..., $b_{10}$) outside the current set ($b_i \notin H$).

Eleven strategies were constructed: one - remaining with the current set of stands, and ten strategies - replacing a with $b_i$ in the set of stands. Next, payoffs were calculated for each player, and a payoff matrix of dimensions (11 x 11 x 11 x 11 x 11) was constructed. Nash equilibrium were then selected from this payoff matrix. Changes were made only when the strategy corresponded to the diagonal of the matrix, meaning the choice of the replacement strategy was a Nash strategy for each player. This simulation was repeated 100 000 times to obtain the final set $H_N$. The game in this format was played 100 times without changing the initial set.

### 3. Results

The statistics of the basic harvesting strategy $H_0$ is presented in the Table 2. The harvesting area are mainly fresh coniferous (>60%) and deciduous (>20%) forests, with protective functions (>90%). The main harvested species are pine (79% of volume), oak (7%), birch (4%) and alder (3%).

**Table 2.** The statistics of the initial set $H_0$ (forest stands initially selected for harvesting)

| Parameter | Value | Harvested species | Mean age | Volume [$10^3 m^3$] |
|---|---|---|---|---|
| **Habitat type** | Coniferous: <br> - fresh: 63.9 % <br> - dry: 1.8 % <br> - wet: 1.0 % <br> Deciduous: <br> - fresh: 21.1 % <br> - wet: 4.3 % <br> Alder forest: 3.2% <br> Riparian forest: 4.7% | **Pine** | $80.0 \pm 32.3$ | 453.3 |

| | | Harvested species | Mean age | Volume [m³] |
|---|---|---|---|---|
| **Forest function** | Protective – 91.7% Economic – 8.3% | **Oak** | 90.9 ± 38.2 | 42.6 |
| | | **Birch** | 56.8 ± 25.4 | 23.0 |
| | | **Alder** | 72.1 ± 32.0 | 15.8 |
| | | **Poplar** | 55.8 ± 25.5 | 7.8 |
| | | **Beech** | 90.1 ± 33.3 | 7.3 |
| | | **Other** | | 21.0 |

The statistics of the strategy $H_N$ is presented in the Table 3. The harvesting area are mainly fresh coniferous (>65%) and deciduous (>20%) forests, with protective functions (>80%). The main harvested species are pine (78% of volume), oak (7%), alder (5%) and birch (2%).

**Table 2.** The statistics of the final set $H_N$ (forest stands selected for harvesting in Nash equilibrium)

| Parameter | Value | Harvested species | Mean age | Volume [m³] |
|---|---|---|---|---|
| **Habitat type** | Coniferous: - fresh: 65.1% - dry: 1.0% - wet: 0.8% Deciduous: - fresh: 20.3 % - wet: 5.2 % Alder forest: 4.4% Riparian forest: 3.2% | **Pine** | 83.1 ± 28.8 | 448.2 |
| **Forest function** | Protective – 68.2% Economic – 31.8% | **Oak** | 103.7 ± 38.6 | 42.6 |
| | | **Alder** | 76.9 ± 30.9 | 26.0 |
| | | **Birch** | 63.1 ± 22.9 | 12.3 |
| | | **Poplar** | 60.7 ± 13.8 | 10.4 |
| | | **Beech** | 104.1 ± 40.8 | 7.5 |
| | | **Other** | | 24.0 |

The payoff values for all five players in the basic harvesting strategy $H_0$:

$$P = \begin{bmatrix} -877.1\ [\text{kt}] \\ -17.8\ [\%] \\ +570\ 874\ [m^3] \\ -6.88\ [-] \\ -2393\ [ha] \end{bmatrix}.$$

The payoff values for all five players in the Nash equilibrium harvesting strategy $H_N$:

$$P = \begin{bmatrix} -667.6 \ [\text{kt}] \\ -10.3 \ [\%] \\ +570\,980 \ [m^3] \\ -6.23 \ [-] \\ -1768 \ [ha] \end{bmatrix}.$$

During the game, the indicators of achieving individual goals evolve, converging towards the Nash-optimal strategy. The figure 1 illustrates the convergence rate of the indicators. The indicators on the graph are normalized to their initial values.

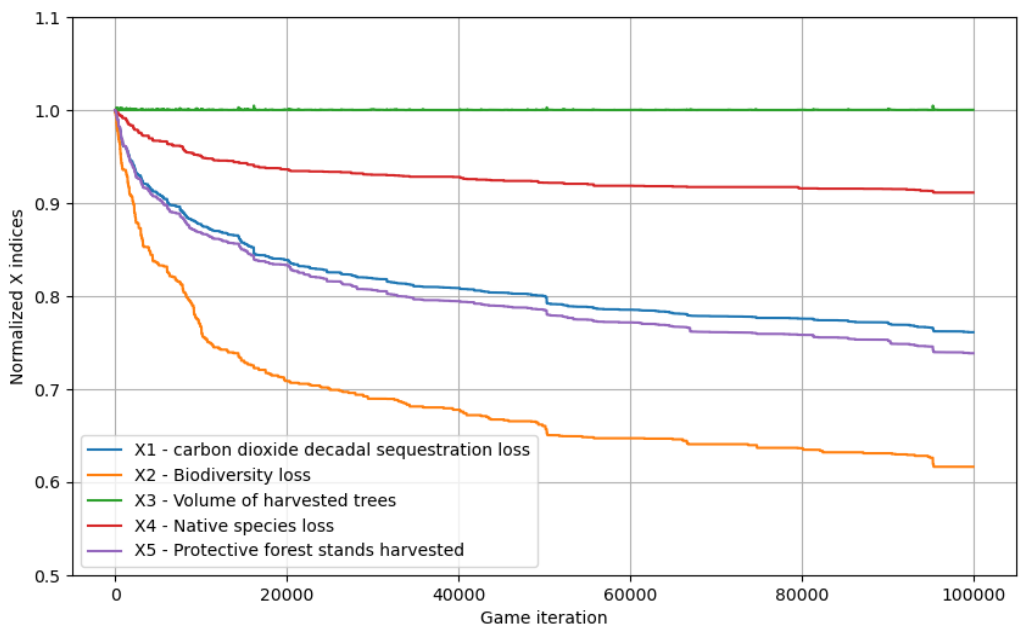

**Figure 1.** Evolution of indicators during a game. Indices are normalized to theirs original values. The assumption was: volume of harvested trees should not decrease, all other indices should decrease.

In the Figure 2, histograms of the age of felled trees categorized by their species are presented.

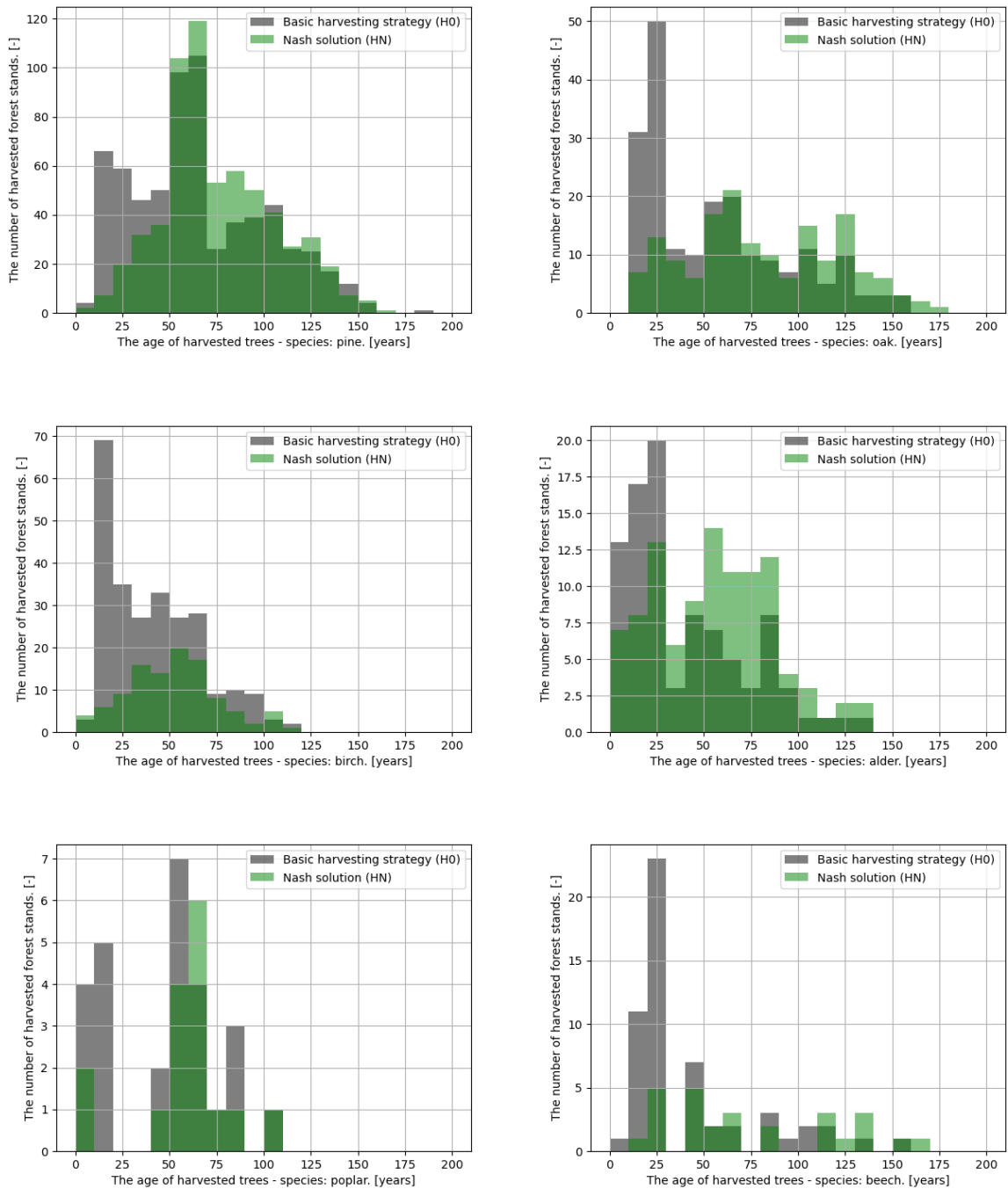

**Figure 2.** Histograms of the age of harvested trees in the basic strategy (grey) and in the Nash equation solution (green). In Nash strategies, one can observe the avoidance of harvesting young trees (< ~50 years old).

## 4. Discussion

In Nash strategies, trees in the youngest age class are removed from harvested forest stands (see: Figure 2). For pines, trees <50 years old are removed, while for oaks, poplars, beeches, and alders, trees <25 years old are removed. In some species, the loss of harvested biomass in the youngest stage is compensated by harvesting additional biomass from older trees: pines aged 75-100 years and alders over 50 years old. This is a sensible solution because tree annual growth rates are increasing during this time, resulting in a higher opportunity cost than the gain from logging (see: Figures 5-10 in Kotlarz and Bejger, 2024, Afzal and Akhtar

2013, Jo and Park 2017). By removing trees at a young age, we miss out on the opportunity for maximum $CO_2$ sequestration. Instead, trees that no longer contribute significantly to $CO_2$ sequestration and are approaching their maximum size are removed. Therefore, despite harvesting the same amount of wood in Nash equilibrium as in the initial strategy, the $CO_2$ sequestration loss is reduced by approximately 25% over a 10-year period (by around 200,000 tons).

No significant changes were observed in the type of habitats designated for harvesting. Fresh habitats still predominate, with a similar distribution of forests and woodlands (approximately 65% woodlands, 25% deciduous forests, and around 10% wetlands). The exchange of forest stands during the game primarily occurred within the main types of forests. However, due to one player maximizing their payoff by reducing harvesting from protective stands, the proportion of forest stands in protective habitats decreased by 25% (625 ha).

Between the initial strategies and the Nash equilibrium, slight differences were observed in the species composition of harvested wood. While the volumes of dominant species (pine and oak) remained unchanged, there was a decrease in the proportion of harvested birch wood (by 10,000 m$^3$), and an increase in the share of alder wood (also by approximately 10,000 m$^3$), representing the most significant change at approximately 1.5% of the harvested wood volume. Additionally, there was a notable shift in the average age of harvested trees: oak experienced a 12-year increase, beech saw a 14-year increase, and the remaining species also showed an increase in age at harvesting by approximately 5 years.

The obtained result indicates the potential for optimizing the selection process of forest stands designated for harvesting. This optimization aims to improve the performance indicators of all forestry management goals in Poland, even those seemingly conflicting. While the simulation was based on illustrative and highly general indicators, its outcomes suggest the need for further exploration and refinement in this area.

## 5. Conclusions

In conclusion, the study sheds light on the potential benefits of applying Nash equilibrium strategies in optimizing forest harvesting decisions. By analyzing the evolution of various indicators throughout the game and observing the resulting changes in harvested forest stands, valuable insights have been gained regarding the balance between different forestry management goals. The findings underscore the importance of considering multiple objectives and trade-offs in forestry decision-making processes. Moving forward, further research and refinement of the optimization approach based on the Nash equilibrium concept could lead to more effective and sustainable forest management practices.